\documentclass[structabstract]{aa}  

\usepackage{natbib}
 \bibpunct{(}{)}{;}{a}{}{,} 
 
\usepackage{graphicx}

\usepackage{txfonts}
\def\kms{km s$^{-1}$\space}
\def\kmsno{km s$^{-1}$}
\def\micron{$\mu$m\space}

\def\micronno{$\mu$m}
\def\arcsecno{$^{\prime\prime}$}

\def\deg{$^{\circ}$\space}

\def\degno{$^{\circ}$}
\def\h2{H$_2$}

\def\cii{[C\,{\sc ii}]\space}

\def\ciino{[C\,{\sc ii}]}
\def\nii{[N\,{\sc ii}]\space}
\def\niino{[N\,{\sc ii}]}

\def\hi{H\,{\sc i}\space}

\def\hii{H\,{\sc ii}\space}

\def\12co{$^{12}$CO}
\def\13co{$^{13}$CO}
\def\c18o{C$^{18}$O}

\def\C+{C$^+$}
\def\N+{N$^+$}
\def\h2{H$_2$}
\def\cm3s{cm$^{-3}$\space}
\def\cm3{cm$^{-3}$}

\begin{document}

       \title{\cii and \nii  from dense ionized regions in the Galaxy}
\titlerunning{\cii and \nii from dense ionized regions}

\authorrunning{W. D. Langer et al.}

   \author{W. D. Langer
             \inst{1},
 	 P. F. Goldsmith
                  \inst{1}, 
                \and 
        	 J. L. Pineda
	 \inst{1} 
                        }
      
         
   \institute{Jet Propulsion Laboratory, California Institute of Technology,
              4800 Oak Grove Drive, Pasadena, CA 91109, USA\\
              \email{William.Langer@jpl.nasa.gov}
             }

   \date{Received 20 January 2016; Accepted 18 March 2016}

 

\abstract
{The interstellar  medium (ISM) consists of highly ionized and neutral atomic, as well as molecular, components. Knowledge of their  distribution is important for tracing the structure and lifecycle of the ISM.}  
 {To determine the properties of the highly ionized gas  and neutral weakly ionized gas in the Galaxy traced by the fine-structure lines of ionized nitrogen, \niino, and ionized carbon, \ciino. }    
 {We utilize observations of the \cii 158 \micron  and \nii 205 \micron fine-structure lines taken with the high spectral resolution  Heterodyne Instrument in the Far-Infrared  (HIFI) on the \textit{Herschel} Space Observatory along ten lines of sight towards the inner Galaxy to analyze the ionized ISM.  The \nii emission can be used to estimate the contribution of the highly ionized gas to the \cii emission and separate the contributions from highly ionized and weakly ionized neutral gas.} 
 {We find that \nii has strong emission in distinct spectral features along all lines of sight associated with strong  \cii emission.  The \nii arises  from moderate density extended  \hii regions or ionized boundary layers of clouds.  Comparison of the \nii and \cii spectra in 31 separate kinematic features shows that many of  the \cii spectra are affected by absorption from low excitation gas associated with molecular clouds, sometimes strongly so.  The apparent fraction of the \cii associated with the \nii gas is unrealistically large in many cases,  most likely due to the reduction of  \cii by absorption.  In a few cases the foreground absorption can  be  modeled to determine the true source intensity.  In these sources we find that the foreground absorbing gas layer has \C+ column densities of order 10$^{18}$ cm$^{-2}$. }  
 {\cii emission arising from strong sources of \nii emission is frequently absorbed by low excitation foreground gas complicating the interpretation of the properties of the ionized and neutral gas components that give rise to \cii emission. } 
{} 

\maketitle



\section{Introduction}
\label{sec:introduction}
The structure and dynamics of the neutral components of the interstellar medium (ISM) are readily traced with spectrally (velocity) resolved ground based observations of \hi 21--cm radiation and the rotational lines of molecular species such as CO and CS.  The ionized gas can be traced by radio recombination lines and H$\alpha$ emission, but observations of the potentially very useful diagnostic lines of \C+ and \N+ has been difficult because their strong fine structure far-infrared emission lines, \cii and \niino, are absorbed by the Earth's atmosphere.

Comparison of \nii with \cii allows one, in principle, to separate highly ionized regions (H$^+$, C$^+$, N$^+$) from neutral weakly ionized regions (H, C$^+$, N) because nitrogen has a higher ionization potential (14.5 eV) than does hydrogen (13.6 eV) so that it is atomic, N, where hydrogen is atomic or molecular, and ionized, N$^+$, where hydrogen is nearly completely ionized.  N$^+$ is produced  by EUV or X-ray photoionization, electron collisional ionization, or proton charge exchange \citep{Langer2015N}.  In contrast, carbon has an ionization potential (11.1 eV) less than that of hydrogen so \C+ arises from all regions suffused with UV.   \cii is thus produced in highly ionized regions, such as the Warm Ionized Medium (WIM), compact and extended \hii regions, and dense ionized boundary layers (IBL) of clouds, as well as in neutral regions which are only weakly ionized (fractional ionization $\lesssim$10$^{-2}$) such as  \hi clouds, PDRs, and CO--dark H$_2$ clouds, whereas \nii arises only from highly ionized regions. 

The COBE FIRAS survey of \cii at 158 \micron and \nii at 122 and 205 \micron in the Milky Way \citep{Bennett1994}  showed that their emission was pervasive, with \cii the strongest far-infrared line, followed by \nii at about one-fifth of the intensity of \ciino.  COBE thus established the importance of \cii and \nii as diagnostic tracers of the ISM where carbon and nitrogen are ionized \citep{Bennett1994,Fixsen1999}.  However, COBE had a 7\deg beam and velocity resolution $\sim$1000 \kms and so could not provide details of the distribution of \C+  and \N+ in the Milky Way, locate the  emission along the line of sight (LOS), or explore the dynamics of the  gas traced by \cii and \niino.   The Heterodyne Instrument in the Far-Infrared \cite[HIFI,][]{degraauw2010} on the {\it Herschel} Space Observatory \citep{pilbratt2010} provided the first opportunity to obtain a large sample of velocity resolved \cii and \nii in the Galaxy in emission \citep{Langer2010,Goldsmith2015} and absorption \citep{Persson2014,Gerin2015}. The determination of the properties of the ionized gas in emission and absorption probe higher and lower density phases of the ISM, respectively.

The  {\it Herschel} Space Observatory Open Time Key Programme, Galactic Observations of Terahertz C+ (GOT C+) conducted a sparse survey of velocity resolved \cii at 157.740 \micron (1900.537 GHz) along over 500 LOS throughout the disk \citep{Langer2010,Pineda2013,Langer2014}. Recently \cite{Goldsmith2015} reported on a follow up survey in \nii of 149 LOS observed in \cii by GOT C+.  All 149 LOS were observed with PACS \citep{Poglitsch2010}  in both fine structure transitions of \niino, $^3$P$_2$--$^3$P$_1$ at 121.898 \micron (2459.371 GHz) and $^3$P$_1$--$^3$P$_0$ at 205.178 \micron (1461.134 GHz).  \nii was detected  in 116 LOS at 205 \micron and in 96 of these at 122 \micronno, where line of sight here means the average of the 25 PACS spaxels with an effective angular resolution $\sim$47\arcsecno. 

\cite{Goldsmith2015} used the two \nii transitions observed with PACS, along with a collisional excitation model,  to derive the average line of sight electron densities, $n$(e).  They found  $n$(e) were typically in the range 10 -- 50 cm$^{-3}$ (although a few had larger values with a maximum of 118 cm$^{-3}$ towards the Galactic center) with an average value 29 cm$^{-3}$, much higher than densities typical of the Warm Ionized Medium.  They derived \N+ column densities, $N$(N$^+$), which are generally in the range 10$^{16}$ -- 2$\times$10$^{17}$ cm$^{-2}$, with the largest value 3.4$\times$10$^{17}$ cm$^{-2}$ towards the Galactic center.  They suggest that these moderately high electron densities in the N$^+$ gas arise either in extended  \hii regions, IBLs around molecular clouds, or from very dense fluctuations in the WIM.   

  \cite{Goldsmith2015} also compared the N$^+$ and C$^+$ column densities derived from PACS  \nii and HIFI \cii emission and found that a significant fraction, $\sim$75\%, of the sources had more than 1/3 of the \cii emission arising from the dense ionized gas (see their Figure 22).  Croxall et al. (2016, in preparation;  J. D. Smith, private communication) found a similar result from a PACS study of \nii and \cii in local star--forming galaxies, namely that a significant fraction of the \cii emission arises from the highly ionized medium.

The low spectral (velocity) resolution of PACS precludes knowing where along the line of sight the emission arises or how many components contribute to a measured intensity and thus the results are a weighted average of the emission from ionized gas in the beam.  Towards the ten strongest \cii LOS in the GOT C+ survey \cite{Goldsmith2015} observed the 205 \micron transition of \N+ with {\it Herschel} HIFI.  \nii was detected in all ten LOS and, together with spectrally resolved \ciino, provides a more detailed look at the nature of the sources of \nii and \cii emission.

 In this paper we analyze the conditions of the ionized gas  in 31 kinematic components arising along the ten LOS observed in \nii with HIFI.  We find that the \nii emission arises from both strong and weak \cii components and that there is significant variation in the \cii to \nii ratio among components and within the line profile.   Furthermore, from comparison of \nii and \cii spectral line shapes, we find that the \cii in several sources is strongly absorbed by lower excitation foreground gas and many of the other sources show at least some trace of \cii absorption. The  \cii absorption, especially in the deep absorption cases, complicates the interpretation of the physical conditions from \ciino.

This paper is organized as follows.  In Section 2 we present the \nii and  \cii results.  Section 3 discusses the properties of gas traced by \cii and \niino, and Section 4 summarizes the results.



\section{Results}
\label{sec:results}
The ten lines of sight observed in \nii 205 \micron and \cii 158 \micron with HIFI are listed in Table 4 of \cite{Goldsmith2015} where their column 1 gives the GOT C+ LOS label and column 2 the Galactic longitude (all sources were observed at $b = $0\degno).  We will use the GOT C+ LOS label here which is in the form Gxxx.x+y.y where the first term gives the Galactic longitude and the second the Galactic latitude, both rounded off to one decimal place. The data reduction of \nii and \cii is described in \cite{Goldsmith2015} and \cite{Pineda2013}, respectively, with the exception that \cii towards the Galactic Center, $(l,b)$ =(0\degno,0\degno),  was reprocessed using HIPE--13 and a different baseline matching routine used to combine observations made with two LO settings.  A baseline matching routine is necessary because the spectrum towards the Galactic Center is so broad that two LO settings are required to fit the entire velocity dispersion of \cii emission. In addition, even with two LO settings, there is only baseline over a narrow band at each end. This narrow baseline band results in some uncertainty in the baseline offset as can be seen in the slight differences in the spectrum offset here and \cite{Goldsmith2015}. 

\begin{figure*}[!ht]
 \centering
   \includegraphics[width=15.5cm]{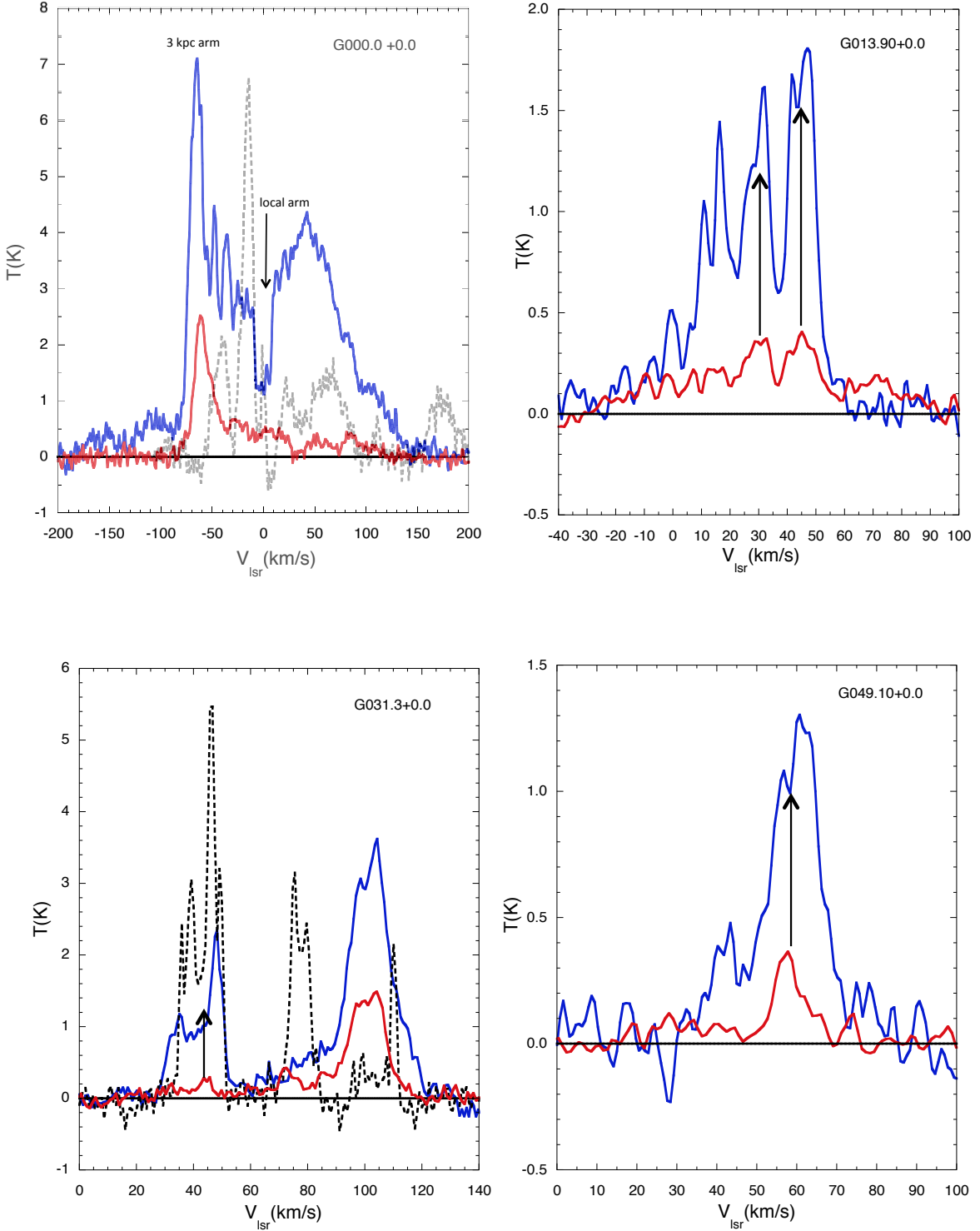}
      \caption{Main beam temperature versus velocity for \cii (blue solid spectra) and \nii (red solid spectra) taken with HIFI -- see Table~\ref{tab:Table_1}  for some of the line parameters, and $^{13}$CO(1-0) (black dashed line) from \cite{Pineda2013}. Each line of sight is labeled with the GOT C+ LOS label. Some features that are indicative of \cii absorption are highlighted with arrows.}
         \label{fig:fig1_CII_NII_13CO}       
  \end{figure*}

\begin{figure*}[!ht]
 \centering
                \includegraphics[width=15.5cm]{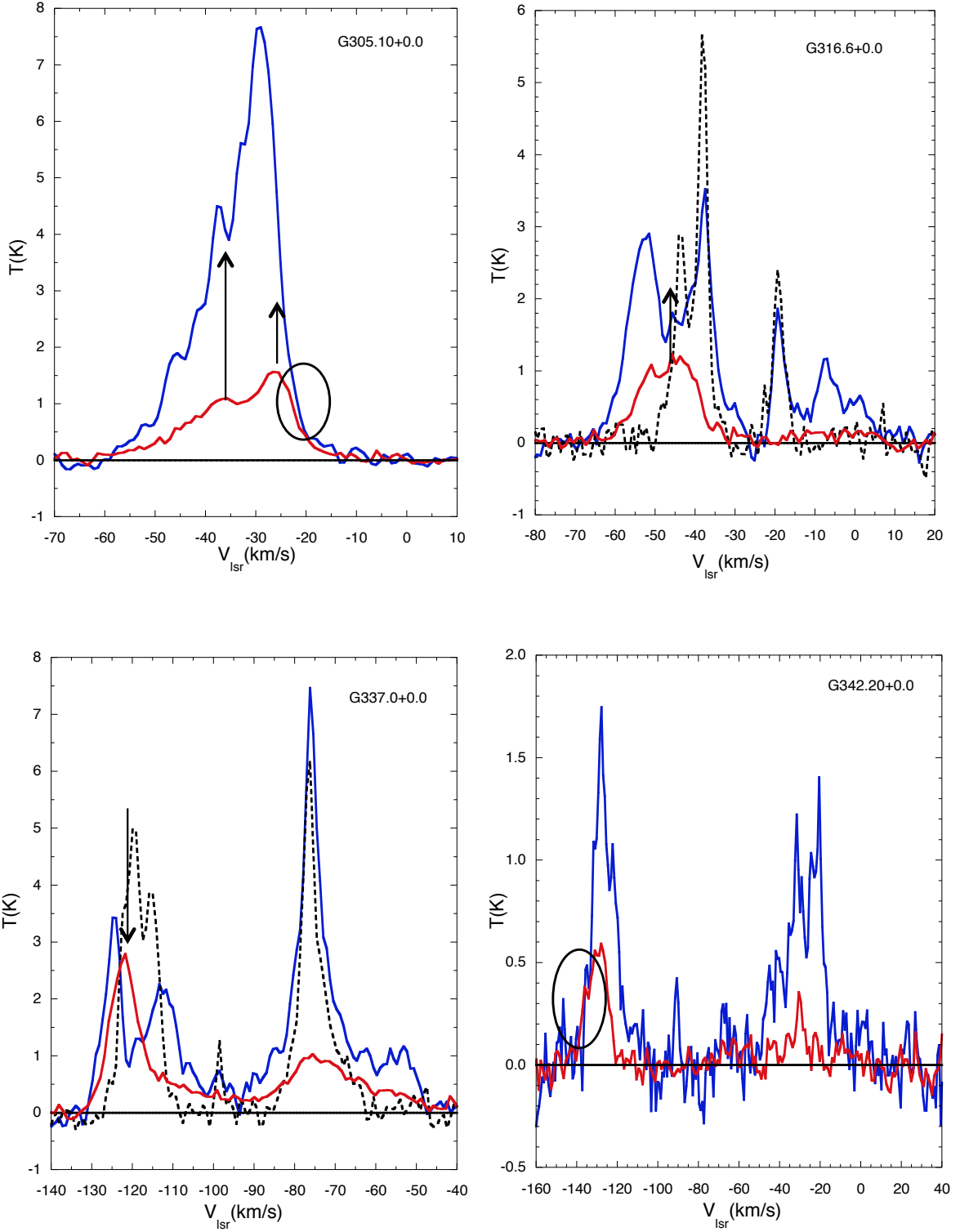}
 \caption{Main beam temperature versus velocity for \cii (blue solid spectra) and \nii (red solid spectra) taken with HIFI -- see Table~\ref{tab:Table_1}  for some of the line parameters, and   $^{13}$CO(1-0) (black dashed line) from \cite{Pineda2013}. Each line of sight is labeled with the GOT C+ LOS label.   Some features that are indicative of \cii absorption are highlighted with arrows and the shoulders of lines where \nii is almost as strong as \cii are highlighted with ellipses.}
         \label{fig:fig2_CII_NII_13CO}       
\end{figure*}

\begin{figure*}[!ht]
 \centering
                \includegraphics[width=15.5cm]{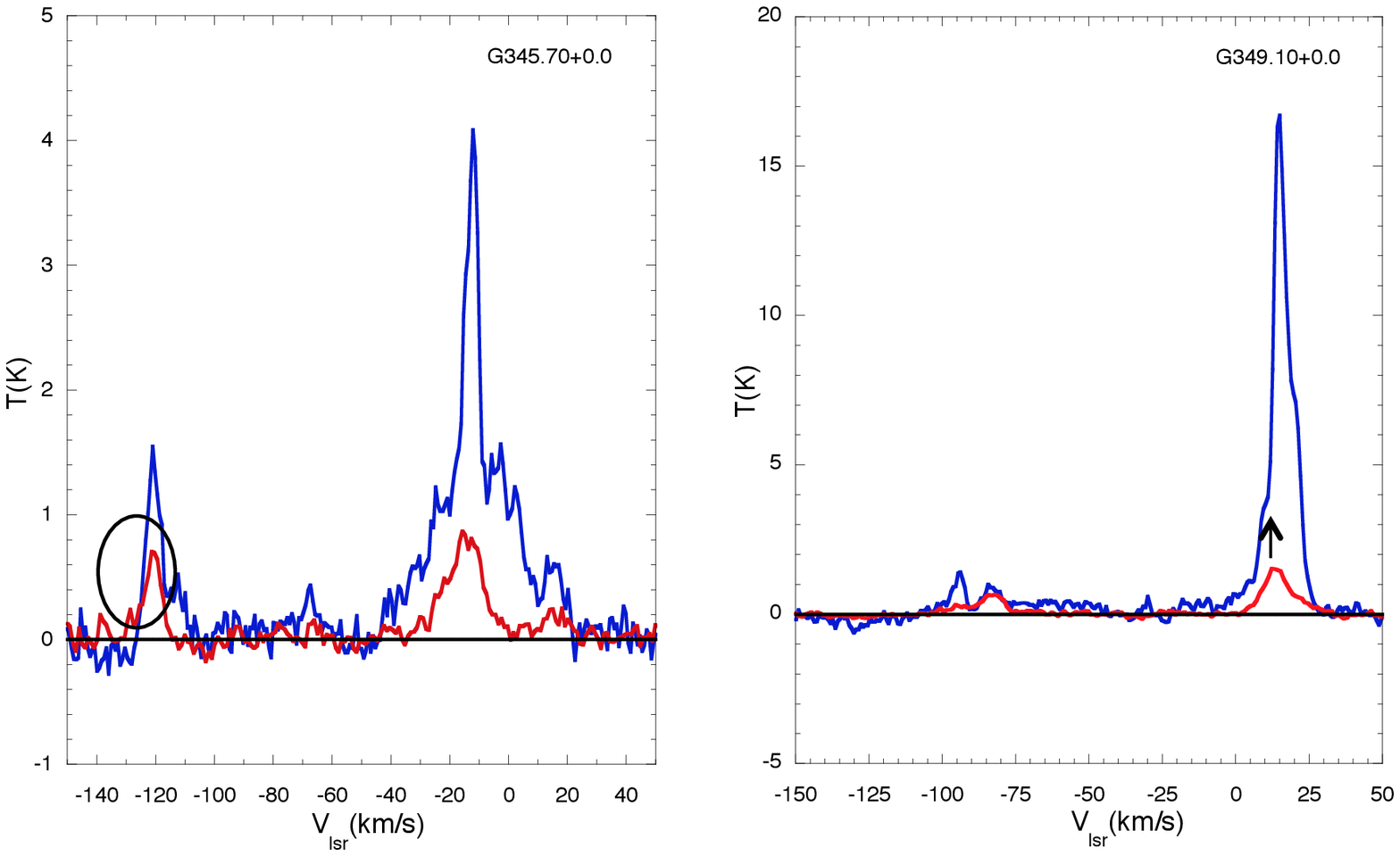}
         \caption{Same as Figure~\ref{fig:fig2_CII_NII_13CO}.}
         \label{fig:fig3_CII_NII_13CO}       
  \end{figure*}

The \cii and \nii spectra for the ten sources are shown in Figures~\ref{fig:fig1_CII_NII_13CO} to \ref{fig:fig3_CII_NII_13CO}. We also include the corresponding $^{13}$CO(1 - 0) spectrum   \cite[][]{Pineda2013} for four lines of sight to guide the identification of the features.  The \13co spectra are sometimes useful to distinguish contributions from the different LOS components where \cii and \nii are strongly blended and to identify dense molecular gas associated with the \nii and \cii sources.  The spectrum towards the Galactic Center is shown in Figure~\ref{fig:fig1_CII_NII_13CO} and   we have indicated on the figure the velocity corresponding to the local and 3 kpc arms.  It can be seen that the strong \nii feature near -60 \kms probably arises from this arm and that the \cii emission from the Galactic center is strongly absorbed by gas likely associated with the local arm.   

Several characteristics of \nii and \cii emission stand out in Figures~\ref{fig:fig1_CII_NII_13CO} to \ref{fig:fig3_CII_NII_13CO}.  First, \nii emission is generally detected wherever there is \ciino, although there is significant variation in the ratio of \cii to \nii intensity as a function of velocity.  Second,  \cii appears to be absorbed or self-absorbed in many components, and strongly so in a few sources.    \cii absorption is very clear, for example, in G316.6+0.0 and G337.0+0.0 where the peak in \nii emission is in or very near a deep minimum in \cii emission.  In some lines of sight, such as G013.9+0.0 and G049.1+0.0, the peaks in \nii align with a smaller dip in the \cii emission (indicated by arrows).  In other lines of sight the evidence is more suggestive than definitive, for example some \nii peaks align with the shoulder in the \cii emission, as in G305.1+0.0, G349.1+0.0, and G342.2+0.0 (indicated by arrows).  Finally,  in G305.1+0.0, G342.2+0.0, and G345.7+0.0 \nii emission is about equal to that of \cii in the shoulders  (highlighted by ellipses). This relationship could be a result of absorption of \cii but could also be due to a velocity shift of the N$^+$ gas layer with respect to the C$^+$ gas in the neutral PDRs.  Third, most of the \nii lines are broad (of order 10 to 20 \kmsno) compared to the velocity dispersion typical for the associated dense molecular clouds.  In some of these lines the corresponding \13co spectra is composed of a couple of narrower components (see for example G316.6+0.0) so the broad \nii and \cii may be due to blending of a few cloud components.  Fourth, \13co is not associated with the full range of \cii emission;  there are \13co spectral features with no detectable \cii and \nii and vice-versa. For example, in G031.3+0.0 there is strong \13co emission near 75 \kms with weak \cii and  \niino, while near 100 \kms the \13co is weak while \cii and \nii are strong. For a more detailed discussion of the relationship of \cii and \13co emission see \cite{Pineda2013} and \cite{Langer2014}. 

Our goal in this paper is to understand the relative contributions of different ISM components to the \cii intensity because of its importance in determining the Galactic \cii distribution and its role in measuring the star formation rate \cite[see][]{Pineda2013,Pineda2014}. Our approach is to use the \nii lines to calculate the expected  \cii contribution from the highly ionized gas and then subtract it from the observed \cii to determine its contribution from the neutral gas.  However, the  absorption in \cii makes determination of  the relative contributions of \cii from highly ionized and neutral gas problematic. In this Section we outline the procedure for calculating the \cii contribution from ionized and neutral components neglecting absorption.   In Section~\ref{sec:discussion}  we will address the correction for absorption.  
  
To calculate the contribution of \C+ from the ionized \N+ region to the measured \cii intensity we need to transform the \nii intensity to the expected \cii intensity.   To derive the relationship of \cii to \nii in  a highly ionized regime we start with the formula for the relationship between intensity and column density for an optically thin species  \citep{Goldsmith2012},
  
\begin{equation}
I_{ul}=\int T_{ul}d{\rm v} = \frac{h c^3}{8\pi k \nu_{ul}^2} A_{ul} N_u
\label{eqn:I_vs_N}
\end{equation} 

\noindent    where $T_{ul}$ is the antenna temperature of the upper ($u$) to lower ($l$) transition, $\nu_{ul}$ is the transition frequency, $A_{ul}$ the Einstein $A$--coefficient, and $N_u$ the column density of the upper level. In a uniform medium we can simplify Equation~\ref{eqn:I_vs_N} by setting $N_{u}= f_u n(X)L$ where $n(X)$ is the density of the emitting species, $X$, $L$ the path length of the emission region, and $f_u$ the fractional population of the upper level, to yield,

\begin{equation}
I_{ul}=\int T_{ul}d{\rm v} = \frac{h c^3}{8\pi k \nu_{ul}^2} A_{ul} f_u n(X)L
\label{eqn:I_vs_N_uniform}
\end{equation} 

\noindent Therefore the ratio of the \cii 158 \micron to \nii 205 \micron  intensity from the \N+ gas, $I_{ion}$(\ciino),  in a fully ionized region,  is given by

\begin{equation}
\frac{I_{ion}([{\rm C\,II]})}{I_{ion}([{\rm N\,II]})}=  \frac{\nu_{10}^2 A_{3/2,1/2} }{\nu_{3/2,1/2}^2 A_{1,0} } \frac{f_{3/2}n(C^+)}{ f_{1}n(N^+)}\,,
 \label{eqn:NII_CII_3}
\end{equation}

\noindent and the contribution of \cii emission from the N$^+$ region is related to the   corresponding  \nii emission by 

\begin{equation}
I_{ion}([{\rm C\,II]})= 0.675 \frac{f_{3/2}({\rm C^+})}{f_{1}({\rm N^+})}\frac{x({\rm C^+})}{x({\rm N^+})} I_{ion}([{\rm N\,II}])\,,
 \label{eqn:NII_CII_4}
\end{equation}

\noindent where $x({\rm C^+})$ and $x({\rm N^+})$ are the fractional abundances of \C+ and N$^+$, respectively, and  $A_{3/2,1/2} \nu_{1,0}^2/A_{1,0} \nu_{3/2,1/2}^2$ = 0.675.  In highly ionized regions Equation~\ref{eqn:NII_CII_4}  is a function of kinetic temperature, $T_k$, and $n$(e) only through the ratio of the fractional populations of the upper levels of \cii ($^2P_{3/2}$) and \nii 205 \micron ($^3P_1$).  We will use the values of $n$(e) derived from PACS data \citep{Goldsmith2015} and  adopt $T_k$= 8000K, characteristic of highly ionized regions, to calculate the ratio of these fractional populations.  However, the exact values of $n$(e) and $T_k$ are not critical because the ratio of the level populations is relatively insensitive to density and temperature as their collisional rate coefficients and critical densities, $n_{cr}$(e), are comparable \cite[see][]{Goldsmith2012,Goldsmith2015}.  For example, over the density range of interest,  $n$(e) = 5 to 150 cm$^{-3}$,  $f_{3/2}/f_1 \sim$1.38  within $\pm$12\%.   The fractional emission derived for \cii depends on the ratio $x$(C$^+$)/$x$(N$^+$). Here we adopt the value 2.9 from \cite{Goldsmith2015}, however the abundance ratio could be lower either because the elemental gas phase carbon to nitrogen ratio is smaller or ionization processes, such as photoionization near X-ray sources, could selectively change the ratio \citep{Langer2015X}. 

The total observed \cii emission is the sum of the ionized, $I_{ion}($\ciino$)$, and neutral,  $I_{neut}($\ciino$)$, ISM components,

\begin{equation}
 I_{tot}([{\rm C\,II}]) =  I_{ion}([{\rm C\,II}]) +  I_{neut}([{\rm C\,II}]),
\end{equation}

\noindent and, in principle, the neutral component can be derived by subtracting the ionized component from the total observed \cii emission.  

 We evaluate the contribution of the ionized gas to the \cii emission for each component by determining $I($\niino$)$ from the HIFI spectra and solving Equation~\ref{eqn:NII_CII_4} using $n$(e)  from \cite{Goldsmith2015} for each LOS.  For many of the \nii spectra the components are well defined with a nearly Gaussian shape, however, some features are a blend of multiple components and for these we make an educated guess of the velocity range of the \nii component.  We have identified 31 kinematic  \nii components with $I($\niino$)$ signal-to-noise $\gtrsim$ 5 (all corresponding \cii components have much higher signal-to-noise).  The line parameters for the 31 components analyzed here are given in Table~\ref{tab:Table_1}, where we list in columns 1 to 9 for each LOS the velocity range of interest, the corresponding intensities of \cii 158 \micron and \nii 205 \micronno, the $V_{lsr}$ and $\Delta$v(\niino) (the full width half maximum)   for \nii and $\Delta$v(\ciino) for \cii wherever we could fit a Gaussian to the blended lines, and the average values of $n$(e) derived by \cite{Goldsmith2015} from PACS \nii 122 and 205 \micron lines. 

In general the linewidths derived for \cii agree with those of \nii given that it is difficult to derive an unambiguous fit for spectra with multiple blended components.  In a few cases the linewidths disagree by more than a few \kms  which is generally due to closely blended features where it is not possible to fit them separately. Finally, we do not derive the linewidths of  \cii lines that are significantly absorbed in the line center as the fit  has little relevance for comparison to the \nii linewidths.  However, as discussed below we will use the \nii lines in these cases to derive the likely \cii source linewidth and intensity.  In   column 10 we list the derived fraction of \cii emission that arises from the N$^+$ region, $I_{ion}$(\ciino)/$I_{obs}$(\ciino), where $I_{obs}$(\ciino) is the total observed \cii emission.  In   column 11 we give this ratio for the few positions where we can correct \cii for absorption, $I_{ion}$(\ciino)/$I_{tot}$(\ciino), as will be discussed below.  In addition to the parameters for the individual spectral components, we also provide the results of integrating across the full spectrum in Table~\ref{tab:Table_1} for comparison to what might be expected from an instrument, such as PACS, that does not resolve the individual components.

\begin{figure}
 \centering
                \includegraphics[width=8cm]{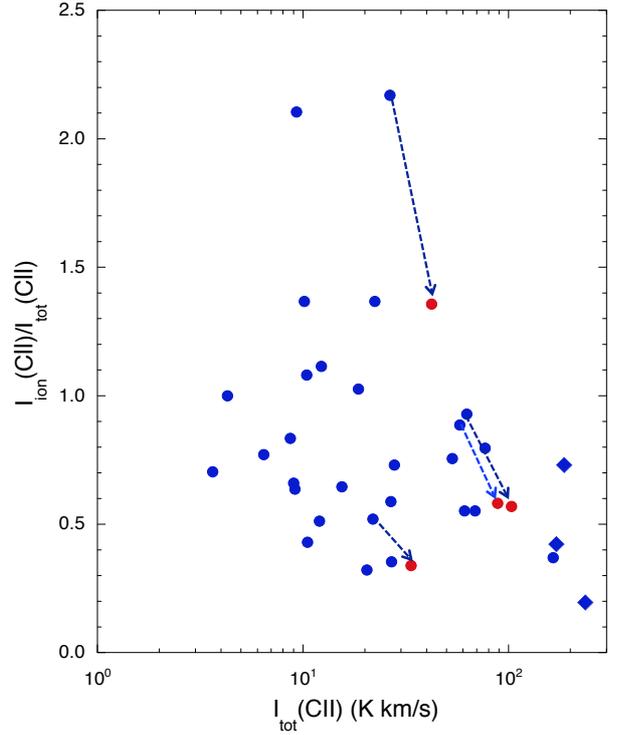}
      \caption{\cii intensity in the ionized gas, $I_{ion}$(\ciino), calculated from the observed \nii intensity, as a fraction of the total \cii intensity, $I_{tot}$(\ciino),  versus the total \cii intensity for thirty one \nii components (28 Galactic Disk and 3 Galactic center).  The blue filled circles and blue filled diamonds  are Galactic Disk and Galactic center sources, respectively. $I_{tot}$(\ciino) equals the observed \cii intensity, $I_{obs}$(\ciino) (uncorrected for \cii foreground absorption).  The four sources which are corrected for foreground absorption are shown by the red filled circles, where the total \cii intensity, $I_{tot}$(\ciino), is the corrected value to the observed intensity. The dashed blue arrows connect the corresponding uncorrected and corrected solutions.  A ratio greater than 1.0 corresponds to a prediction of  \cii arising from the \nii region greater than that observed.  }
         \label{fig:fig4_CII_i_CII_tot}       
 \end{figure}

\begin{figure}
 \centering
                \includegraphics[width=8cm]{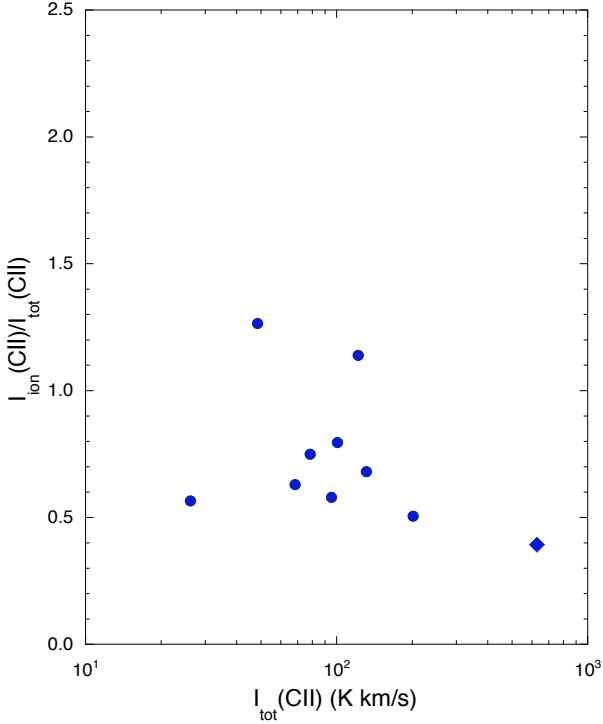}
      \caption{Same as Figure~\ref{fig:fig4_CII_i_CII_tot} but calculated for the total emission within the \cii and \nii bands along each of ten lines of sight   as it would be observed by a low resolution spectrometer.  The blue filled circles and blue filled diamond  are the Galactic Disk sources and the Galactic center source, respectively, where $I_{tot}$(\ciino) equals the observed \cii intensity, $I_{obs}$(\ciino), uncorrected for foreground absorption.  A ratio greater than 1.0 corresponds to a predicted  \cii arising from the \nii region greater than that observed.  }
         \label{fig:fig5_CII_i_CII_tot_LOS}       
 \end{figure}

In  Figure~\ref{fig:fig4_CII_i_CII_tot} we plot the solutions for the fractional \cii intensity arising from the N$^+$ region for all the individual components listed in Table~\ref{tab:Table_1}.  In addition, to simulate the solutions for the unresolved PACS  \nii spectra, we also calculated the total intensity in the HIFI \nii and \cii bands along each line of sight.  These ten fractional intensities are plotted in Figure~\ref{fig:fig5_CII_i_CII_tot_LOS}. For the Galactic Disk (blue filled circles) and Galactic center (blue diamonds) solutions we used the observed \cii intensity for the total intensity, whereas the four red filled circles  represent solutions for \cii emission corrected for absorption (discussed in Section~\ref{sec:discussion}) and for these values we use the total corrected \cii intensity. The ratio of $I_{ion}($\ciino$)$/$I_{tot}($\ciino$)$ must be less than one or else the fractional \cii emission arising from the ionized gas is greater than the total observed \ciino.  

 It can be seen that roughly one quarter of the ratios plotted in Figure~\ref{fig:fig4_CII_i_CII_tot} have $I_{ion}$(\ciino)/$I_{tot}$(\ciino)  greater than unity and, thus, represent an unphysical solution (before correcting for absorption).  In addition a majority of the points have $I_{ion}$(\ciino)/$I_{tot}$(\ciino)  greater than 0.5, which while theoretically possible, is suspicious because it would indicate that the vast majority of  \cii comes primarily from the \nii regions.  However, it is more likely that the observed \cii is not tracing all the \C+ in the ionized and neutral gas due in part to absorption of the \cii emission. The fraction of \cii arising from the ionized gas for the integrated line of sight emission in Figure~\ref{fig:fig5_CII_i_CII_tot_LOS} lie mainly in the range 0.5 to 0.8 in our sample of the disk, with two LOS showing values greater than one, while the Galactic center LOS has the smallest fraction. The fraction for the total LOS emission is generally less than that of the individual components, reflecting the result of averaging regions of differing physical and emission properties.  
 
 A similar result can be inferred from the PACS data in \cite{Goldsmith2015} Figure 22, where they compare the C$^+$ column density $N$(C$^+$) from the GOT C+ data with the N$^+$ column density $N$(N$^+$) derived from PACS.  The \C+ emission that comes from  the N$^+$ region is defined by a line corresponding to $N$(C$^+$)=2.9$N$(N$^+$) on their figure.  Roughly 15\% of the data points fall below that line corresponding to less observed $N$(C$^+$) than predicted from scaling \niino, which likely results from \cii absorption, as seen in our HIFI data for individual spectra. Lines at $N$(C$^+$)=2$\times$2.9$N$(N$^+$) corresponds to half of the \C+ residing in the \nii region.  Roughly half the data points lie in the range of 50\% to 100\% of the \C+ associated with N$^+$, which while possible, is again suspect.  As discussed above, to a good approximation, the ratio of column densities is equal to that of the intensities within 10 to 20 percent for optically thin lines, so the result from PACS holds true for the fraction of \cii emission arising from the \nii region. Although the $N$(C$^+$) to $N$(N$^+$) derived from PACS may be suspect, the  $n$(e) derived by \cite{Goldsmith2015} for the average LOS  should be reasonably good because they depend on the ratio of the \nii 122 and 205 \micron lines, and these lines are less likely to be affected by absorption.  
 
In Section~\ref{sec:discussion} below we show that it is possible  to  correct the fraction of \cii arising from the \nii emission region by accounting for the foreground absorption in four of the sources  in which the \nii and \cii line shapes allow us to estimate the source intensity and correct for absorption.  The foreground absorption is significant and results in at least a 40\% reduction in the observed  \cii arising from the dense ionized regions.

\begin{table*}[!htbp]																		
\caption{\nii and \cii Kinematic Features and Line Properties}
\label{tab:Table_1}														
		\begin{tabular}{lcccccccccc}
\hline																			
 LOS &  \# & $V$ range  & $I$(\ciino)$^a$& $I$(\niino)$^a$ & $V_{lsr}$(\niino)$^b$ & $\Delta$v(\niino)$^b$&  $\Delta$v(\ciino)$^b$ & $n$(e)$^c$& $I_{ion}$(\ciino)/& $I_{ion}$(\ciino)/ \\ 
label	  &  	&   & &  &  & &  &  & $I_{obs}$(\ciino)  & $I_{tot}$(\ciino) \\
  	  &  	& (\kmsno)  & (K \kmsno) & (K \kmsno) & (\kmsno) & (\kmsno) &  (\kmsno) & (cm$^{-3}$) & observed & $\tau^d_{corrected}$ \\

    \hline
  \hline 
G000.0+0	& 1 &  -90$\rightarrow$-35 & 186.8 &  54.0 & -60.3 & 16.8 & 18.7 &  118 & 0.73 & - \\
  			& 2 &  -35$\rightarrow$30 & 171.0 & 28.6 &- & -& - &118 & 0.42 &-\\
    			& 3 & 30$\rightarrow$140  &  236.6 & 18.3 & -&- & - &118 & 0.20 & -\\
			& total & -200 $\rightarrow$200  &  629.1 & 97.9 & -&- & -& 118 &  0.39& -\\
			\hline
G013.9+0  	& 1 &  -5$\rightarrow$5   & 4.3  & 1.7 & - & -&  -& 27 & 1.00 & -\\
    			& 2 &   5$\rightarrow$15  & 9.1 & 2.4 & -& -&  -& 27 & 0.64 & -\\
			& 3 &   15$\rightarrow$25  & 12.0  &2.5 & 30.4 & 15.9 &12.9  & 27 & 0.51 & -\\
			& 4 &   25$\rightarrow$35  &15.5 &4.1 & 45.1 & 14.4 & 11.3 & 27 & 0.64 & -\\
			& 5 &   35$\rightarrow$60  & 26.8  &  6.4 & - & - & - & 27  & 0.59 & -\\
			& total & -50 $\rightarrow$60  &  68.3 & 17.4 & -&- & - &  27 &  0.63 & -\\
\hline
G031.3+0 	&  1&   25$\rightarrow$60  & 27.0 & 3.9 & -&- & - & 31 &  0.35 & -\\
			& 2 &   60$\rightarrow$85  & 10.2 & 5.7 & - & -&  - & 31 &  1.37 & -\\
    			& 3 &   85$\rightarrow$120  & 62.7 & 23.9 & 100.8&15.5  & 18.4 & 31 &0.93 & 0.57 \\
			& total & 0$\rightarrow$130  &  100.7 & 32.8 & -&- & -& 31 &  0.80 & -\\
\hline
G049.1+0  	&  1 &   30$\rightarrow$45  &  3.6  & 1.0 & -  & - & - & 20 & 0.74 & - \\
			&  2 &   45$\rightarrow$75  &  21.9  & 4.5 & 57.9  & 8.7 & 14.2$^e$ & 20 & 0.55 & 0.34 \\
			& total & 0 $\rightarrow$90  &  26.2 & 5.8 & -&- & -& 20 &  0.57 & -\\
\hline
G305.1+0  	&  1 &  -60$\rightarrow$-40    & 27.8  & 8.5  & -45.5& 10.8$^f$& - & 56 &0.82 & - \\
			& 2 &   -40$\rightarrow$-30   & 61.0  & 14.1 & -36.6 & 8.3$^f$&  - & 56 & 0.62 & -\\
    			& 3 &   -30$\rightarrow$-10  & 53.2 & 16.8 & -26.5 &8.8$^f$ &  - & 56 & 0.84  & - \\
			& total & -80 $\rightarrow$0  &  131.2 & 37.3 & -&- & -& 56 &  0.68 & -\\
\hline
G316.6+0  	&  1 &  -70$\rightarrow$-25   & 58.0 & 20.5 &-46.5  &16.4  & - & 23 & 0.94 &0.58\\
			&  2 &  -25$\rightarrow$15   & 20.5 & 2.6 &- & - &  -& 23 & 0.34 & -\\
			& total & -80 $\rightarrow$20  &  78.4 & 23.4 & -&- & -& 23 &  0.75 & -\\
\hline
G337.0+0  	&  1 &  -135$\rightarrow$-115   & 26.5  & 23.5& -121.8 &9.3 & & 30 & 2.37 & 1.36 \\
			&  2 &  -115$\rightarrow$-95   & 18.6  & 7.8& - & - & -& 30 & 1.12 & - \\
			& 3 &   -95$\rightarrow$-40  &  76.9 & 25.0 & -73.7 & 20.5$^e$ & 7.3  & 30 & 0.87 & - \\
			& total & -140 $\rightarrow$-40  &  121.9 & 56.7 & -&- & -& 30 &  1.14 & -\\
\hline
G342.2+0  	&  1 &   -150$\rightarrow$-110  & 22.3 & 11.5 & -129.9 & 13.0 & 12.5 & 14 & 1.37 & -\\
			& 2 &   -50$\rightarrow$-35  & 6.5  &1.9  &- & -& -& 14 &0.77  & - \\
			& 3& -35$\rightarrow$-26 & 8.7& 2.7& - & -& -& 14 &0.84 &-\\			
			& 4 & -26$\rightarrow$-10 &10.5 & 1.7 &-  & -& -& 14& 0.43&-\\			
			& total & -150 $\rightarrow$-40  &  48.5 & 23.1 & -& - & -& 14 &  1.27& -\\
\hline
G345.7+0  	&  1 &  -130$\rightarrow$-100   & 12.3  & 5.7 & -121.0 & 6.9 & 6.5 & 60 & 1.24 & -\\
			& 2 &   -50$\rightarrow$5  & 68.8 & 15.9 & -14.9 & 14.7$^e$ & 9.0 & 60 & 0.62 & -\\
			& 3 &   5$\rightarrow$25  & 9.0 & 2.5 & - & - & - & 60 &0.74 & - \\
			& total & -130 $\rightarrow$25  &  95.4 & 23.10& - &- & -& 60 &  0.58 & -\\
\hline
G349.1+0 	&  1 &  -110$\rightarrow$-88   & 10.4   & 4.7& -83.1 & 9.4 & 7.7 & 66 & 1.20 & -\\
    			& 2 &  -88$\rightarrow$-75  &  9.3   & 8.2 & - & - &  - & 66 & 2.33 & -\\
			& 3 & -5$\rightarrow$35   &  164.9  & 25.4 & 13.7 & 10.7 & 8.6 & 66 & 0.41 & -\\
			& total & -110 $\rightarrow$35  &  201.6 & 42.3 & -&- & - & 66 &  0.51 & -\\
\hline	
\end{tabular}
\\	
a) The intensities in this table are  derived over a velocity range either  judged by eye, or fixed using a Gaussian wherever \nii spectra could be fit unambiguously, in which case we also list their $V_{lsr}$ and $\Delta$v.  The second column numbers the individual components in order of appearance in the band from low to high velocity.  Most values are rounded off to one decimal place. (b) V$_{lsr}$ and $\Delta$v from a Gaussian fit. (c) Derived by \cite{Goldsmith2015} from PACS \nii 205 \micron and 122 \micron data. (d) See Section~\ref{sec:discussion} and Table~\ref{tab:Table_2} for correction for absorption.   (e) Blend of two closely spaced lines broadens the fit.	 (f) Multigaussian fit to three overlapping lines with distinct local maxima.	
 \end{table*}



\section{Discussion}
\label{sec:discussion}
The results presented in Section~\ref{sec:results} show that the observed  \cii in many kinematic components is probably not tracing all the  material in which it originates.  The most likely scenario to explain the absorption suggested by the spectra in Figures~\ref{fig:fig1_CII_NII_13CO} to \ref{fig:fig3_CII_NII_13CO}  is that emission of \cii from a high excitation region on the far side of a cloud is absorbed by low excitation \cii on the near side, or by a few clouds along the line of sight,   or for the inner Galaxy, where there is near- far-distance ambiguity, that emission from a distant spiral arm is absorbed by the near spiral arm at the same velocity.  The absorbing gas must have an excitation temperature considerably lower than that of the emission region.   In addition to the absorption seen in \cii by comparison with the  \nii spectra shown here, absorption of \cii and \nii has also been seen in several sources against bright star forming regions \citep{Gerin2015,Persson2014} and is attributed to the warm ionized medium for \nii and, mainly, to the cold neutral medium for \ciino.  Absorption in \cii has also been inferred in very bright \hii regions from comparison of \cii with its weaker isotopologue [$^{13}$C\,{\sc ii}]  \citep{Graf2012,Ossenkopf2013,Stutzki2013}.    As discussed above, other factors that could affect the fractional emission derived for \cii are optically thick \cii lines and the adopted value of the ratio of \C+ to N$^+$.  If the ratio $x$(C$^+$)/$x$(N$^+$) is smaller than the value adopted in Section~\ref{sec:results},   for example if the elemental gas phase carbon to nitrogen ratio is lower,  then the intensity $I_{ion}$(\ciino) will be smaller.   

\subsection{Spectral line absorption}
Absorption of \cii and \nii by a low excitation envelope or by foreground gas (both local  or in a near-distance spiral arm) will be different for  \N+ and \C+ because each of these species occupy different volumes along the line of sight.  \C+ will be found in highly ionized regions (the warm ionized medium) and in neutral regions (the warm neutral medium, \hi clouds, and CO-dark \h2 regions), but \N+ is  present essentially only in the warm ionized medium, \hii regions, and IBLs. Furthermore, in the highly ionized regions  \N+  will be less absorbed than \C+ because it has a lower column density due to its lower fractional abundance relative to \C+ (here $x(C^+)/x(N^+) \sim$2.9).  Finally, we might expect different degrees of absorption due to the geometry of the gas and the requirement that absorption occurs mainly from gas that is within the velocity range of the emission region. 

Along the ten lines of sight observed with HIFI the electron densities derived by  \cite{Goldsmith2015}  (see Table~\ref{tab:Table_1}) are high enough to readily excite the upper fine-structure levels of N$^+$ and produce the strong \nii emission observed with HIFI.  For $n({\rm e}) \sim$ 20 - 50 cm$^{-3}$ the excitation temperature for the \nii in the ionized gas is high, $\sim$ 40 - 50 K, and similarly for \ciino.  The \N+ column densities derived by \cite{Goldsmith2015} range from 3 to 19$\times$10$^{16}$ cm$^{-2}$ for the Disk, with an average value $\sim$10$^{17}$ cm$^{-2}$,  and 3.4$\times$10$^{17}$ cm$^{-2}$ for the Galactic center.  We cannot know from the PACS data the actual electron densities of each velocity component or range, but it seems a reasonable starting point to adopt a characteristic density for the brightest  \nii components derived from PACS.  The column density derived from PACS is a total column density along the line of sight and the individual component column densities will be smaller where there are multiple components of comparable intensity, as seen in Figures~\ref{fig:fig1_CII_NII_13CO} to \ref{fig:fig3_CII_NII_13CO}  for G013.9+0.0, G031.3+0.0, G305.1+0.0, G337.0+0, and G345.7+0.0.

\subsubsection{\nii opacity}
\label{sec:NII_opacity}
To determine the potential effect of absorption on the \nii spectral lines we first estimate the maximum opacity for the ground state transition in the emission region using the density and column density solutions from \cite{Goldsmith2015}.  The opacity of the $^3P_1$--$^3P_0$ transition for \nii is given by \cite[][]{Persson2014,Goldsmith2015}

\begin{eqnarray*}
 \tau_{10}(N^+) & = & \frac{g_1A_{10}c^3}{g_0 8\pi \nu_{10}^3} \frac{N_0({\rm N^+})}{10^5\Delta {\rm v}} \\
 & =  & 2.1\times10^{-17} \frac{N_0({\rm N^+})}{\Delta {\rm v}}\,\, {\rm cm^{-2}},
  \label{eqn:NII_tau}
\end{eqnarray*}

\noindent where $\Delta$v is the linewidth in \kms and $N_0($N$^+$) is the column density of \N+ in the $^3P_0$ state in cm$^{-2}$. The fractional population in the  $^3P_0$ state is $\sim$0.5  for the electron densities derived for the \nii emission region.  For a typical column density $N$(N$^+$) $\sim$10$^{17}$ cm$^{-2}$ and  \nii linewidth of $\sim$10 \kmsno, $N_0($N$^+$) $\sim$ 5$\times$10$^{16}$ cm$^{-2}$ and $\tau_{10}($N$^+$) $<$ 0.1.  As the total column densities for \N+ in the emission region \citep{Goldsmith2015} are less than 2$\times$10$^{17}$ cm$^{-2}$ in the Disk and 3.4$\times$10$^{17}$ cm$^{-2}$ towards the center, the \nii opacity in the emission region is small and self-absorption in \nii is negligible.  In the foreground absorption region, where the densities are lower, the fractional population in the ground state is $\sim$1 and it would take a column density $N($N$^+$)$\gtrsim$5$\times$10$^{17}$ cm$^{-2}$ to have any significant absorption.

There are three possible sources of \nii absorption, the extended WIM along the line of sight, a lower density portion of the extended \hii region, and the ionized boundary layer along the front side of the cloud   or near-distance spiral arm.  \cite{Persson2014} detected only two \nii sources in absorption with a ground state column density of absorbing N$^+$ $\sim$1.5$\times$10$^{17}$ cm$^{-2}$ and attributed it to the extended WIM.  They estimated the WIM gas density  to be $n$(H$^+$) $\sim$0.1 cm$^{-3}$.  The absorption detected by \cite{Persson2014} occurs over kpc scales and velocity dispersion $\sim$40 - 50 \kms much larger than that of \nii in emission observed in our sources. Thus absorption by low density WIM is unlikely to be significant in our HIFI sources due to the lack of a large column density for $N_0$(N$^+$) within the velocity range of \nii emission features.  If absorption is due to the dense ionized boundary layer on the front side of the cloud then the total column density $N$(N$^+$) derived by \cite{Goldsmith2015} for disk sources corresponds to  $\tau$ (N$^+$) $\lesssim$0.2 and even smaller in the case of multiple components along the line of sight.  Therefore \nii is only slightly absorbed by gas local to the emission region and this result is consistent with the low level of absorption seen in the \nii spectra in Figures~\ref{fig:fig1_CII_NII_13CO} to \ref{fig:fig3_CII_NII_13CO}. 

\subsubsection{\cii absorption}
\label{sec:CII_absorption}
Whereas there is no, or at most very little, absorption seen in the \nii spectra, there is significant to moderate absorption of \cii in most of the ten LOS.  If we can calculate the \cii intensity and line shape of the source we can determine the true contributions of \cii in the ionized and neutral gas, as well as  determine the opacity of the absorbing layer.   However, determining the properties of the absorbing material and its column density is difficult in most cases because the \cii profiles are either too complex or too blended to generate a unique solution to the emission source line.  

\begin{figure}
 \centering
                \includegraphics[width=8cm]{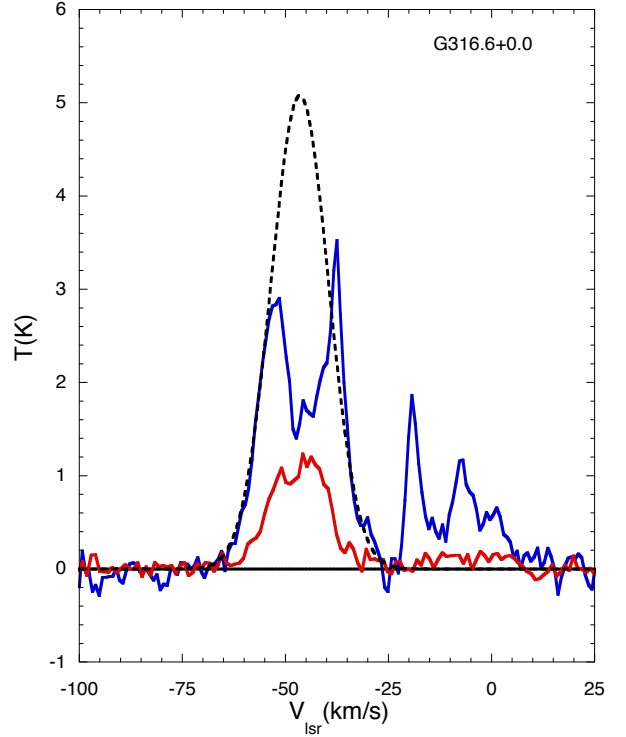}
      \caption{A Gaussian fit (black dash) to the line wings of the $V_{lsr} \sim$ -47 \kms \cii component (blue) in G316.6+0.0 using the $V_{lsr}$ and line width of the Gaussian fit (not shown) to the \nii emission (red). }
         \label{fig:fig6_G316.6_Gauss}       
 \end{figure}

There are four lines of sight for which  it is possible to estimate the profile and strength of the \cii source  by fitting the intensity as a function of velocity in the \cii line wings with a Gaussian, using information from the corresponding \nii spectra as constraints on the \cii solution.  To estimate the true \cii emission line profile that would be observed in the absence of absorption we fit a Gaussian to the wings of the \cii line, adopting a two step process.  First we scale the peak T$_{mb}$(K) of the Gaussian fit to \nii fixing  V$_{lsr}$ and $\Delta$v  until the line wings match those of \cii as well as possible, where the line wings are defined as the regions outside of the FWHM.  This procedure assumes that the line profiles of the \cii and \nii can be represented by similar Gaussians (having comparable $V_{lsr}$ and $\Delta$v), which should be an adequate first approximation   because we find comparable linewidths for \cii and \nii in those cases where the lines can be fit unambiguously, as seen in Table~\ref{tab:Table_1}. We then refine the Gaussian fit to the \cii line wings adjusting the peak temperature, $V_{lsr}$, and $\Delta$v until we minimized the difference across the observed line wings.   In Figure~\ref{fig:fig6_G316.6_Gauss} we show the Gaussian fit representing the source, along with the observed \cii and \nii spectra, for one LOS, G316.6+0.0.  In Table~\ref{tab:Table_2} we list the peak source antenna temperature, $T_s$ and  integrated intensity, $I_{tot}$(\ciino), of the Gaussian fit to the \cii line wings for the four sources. 

We then recalculate the fraction of \cii that arises from the \nii region using the revised total intensity ($I_{tot}$(\ciino) $>$ $I_{obs}$(\ciino)) and the values are  given in Table~\ref{tab:Table_2} along with the uncorrected value from Table~\ref{tab:Table_1}.  Comparison of the fitted \cii emission with that expected from the region producing \nii emission reduces the fraction of \cii from the ionized region by  about  40\% in the four sources.  The revised values for the fractional \cii arising from the ionized regions for these four sources is also plotted separately in Figure~\ref{fig:fig4_CII_i_CII_tot} as red filled circles, an arrow connects the values before and after correction for absorption. It can be seen that three of the four corrected sources have the fractional contributions to \cii from the \nii region reduced to $\lesssim$0.6, but that the source with the highest ratio still has a fraction greater than unity.  There are two possible explanations, either we have underestimated the absorption and/or the \cii lines are optically thick and need to be corrected for it.

\begin{table*}[!htbp]																		
\caption{\cii emission and absorption components}
\label{tab:Table_2}
		
\begin{tabular}{lccccc|cccccc}
\hline	
 & &   & \cii Source &   & &   & & \cii Absorber &  &  &   \\
  & & & Component &  & & & &  Component &   \\															
\hline	
  LOS &  \# & $T_s$  & $I_{tot}$($\tau$) & $I_{ion}$/$I_{obs}$ & $I_{ion}$/$I_{tot}$ & $T_o$  & $V_{lsr}$  & $\Delta$v &  $\tau_0$ & $N$(C$^+$) & $A_v$ \\
  	label   & &(K) &	 (K \kmsno) & observed & corrected & (K) &	(\kmsno) & (\kmsno)   &   & 10$^{18}$ cm$^{-2}$ & mag. \\
  \hline
  \hline 
G031.3  & 3 & 6.3 &  103.5 & 0.93 & 0.56 & 2.9 & 101  & 14.5 &   0.76 & 1.6 & 2.9  \\
G049.1&  2& 3.7   & 33.6 & 0.55 &0.34 & 1.5 &  59 & 6 .7 &   0.89 & 0.9 & 1.6  \\
G316.6  & 1 & 5.1  & 88.5 & 0.94 & 0.58 & 1.4 & -47  & 8.3  &  1.3 & 1.5 & 2.8  \\
G337.0  & 1 & 4.3  & 42.3 & 2.37 & 1.37 & 1.0 & -121  &  4.5   & 1.5 & 0.9 & 1.7  \\  
\hline	
\end{tabular}
\\	
 \end{table*}												

\subsubsection{Foreground opacity}
 Assuming the Gaussian fits to \nii and the line wings of \cii in Section~\ref{sec:CII_absorption} are correct we can calculate the \cii foreground opacity from the projected \cii source intensity, $I_{tot}$(\ciino), and the observed \cii intensity, $I_{obs}$(\ciino), as a function of velocity.   At line center we define the opacity in terms of the reduction in the peak intensity, $\tau_0 = ln(T_s/T_o)$, where $T_s$ is the   corrected peak main beam temperature of the source, and $T_o$ is the observed main beam temperature at $V_{lsr}$.  We neglect \cii re-emission from the foreground absorbing gas as it is likely to be small at low excitation temperatures.  If there were  re-emission then  $\tau_0$ would be underestimated. In Table~\ref{tab:Table_2} we list $T_s$ and $T_o$ for each LOS component along with the $V_{lsr}$ and $\Delta$v of the absorption.  The peak foreground opacity $\tau_o$ is given in column 1.  The column density of the foreground \C+ in the low excitation limit is \cite[Eq. (35),][]{Goldsmith2012},

\begin{equation}
N({\rm C^+}) = 1.3\times 10^{17} \tau({\rm C^+})\Delta{\rm v}\,\, {\rm cm^{-2}}
\label{eqn:tau}
\end{equation}

\noindent where $\Delta$v is in \kmsno.  

We find that $N$(C$^+$) ranges from 0.9 to 1.6 $\times$10$^{18}$ cm$^{-2}$.  The  opacity of the absorbing layer in G316.6+0.0 and G337.0+0.0, $\tau_0 \sim$ 1.3 to 1.5, are the largest of the four sources.  The absorption in these strong sources is consistent with the deep self-absorption in other \cii sources derived from comparison of hyperfine lines of [$^{13}$C\,{\sc ii}] with \cii in a number of bright \cii sources such as NGC2024 \cite[][]{Graf2012,Stutzki2013} and MonR2 \citep{Ossenkopf2013}, and a suggestion of absorption in NGC 3603, Carina North, and NGC 7023 \citep{Ossenkopf2013}. This comparison is consistent with the \cii and \nii emission in G316.6+0.0 and G337.0+0.0 arising from bright extended \hii sources. The remaining two sources G031.3+0.0 and G049.1+0.0 are weaker \cii sources and the corresponding absorbing layers have a lower peak optical depth  $\tau_0 \sim$0.8 .

To compare the thickness of the \C+ absorbing layer with models we need to convert $N$(C$^+$) to visual extinction, $A_v$(mag.).  We assume that the absorbing layer is located in a low density \h2 envelope.  We adopt a fractional abundance $x$(C$^+$) with respect to \h2 of 6$\times$10$^{-4}$ appropriate to the inner Galaxy, $R_{gal} \sim$ 3 - 4 kpc \citep{Pineda2013}, where the \cii and \nii sources are likely located, and the values of $A_v$ are included in the last column of Table~\ref{tab:Table_2}. The estimated thickness of the absorbing layer in the components in two of the \cii sources, G049.1+0.0 and G337.0+0.0, is comparable to those predicted by PDR models of clouds \citep{Visser2009,Glover2010,wolfire2010,Glover2011} in which the atomic hydrogen plus CO-dark \h2 layer has $A_V \sim$ 1  mag. roughly independent of cloud mass, density, and radiation field. Thus absorption in these two \cii sources can probably be explained with standard models of the layer of low density atomic and molecular gas layers.  However, two sources, G031.3+0.0 and G316.6+0.0, have absorbing layers with column densities or extinctions much larger than predicted by the standard PDR models.  One simple explanation could be that a couple of clouds at comparable velocities located in proximity to each other combine to produce the total line of sight absorption.  The bright \cii sources in these ten lines of sight occur in the dense spiral arms where such crowding is likely.    For inner Galaxy sources another possibility is that emission from a distant spiral arm is absorbed by clouds in a near-distance spiral arm at the same velocity. 

\subsection{Carbon to nitrogen abundance}
In addition to absorption of \cii emission by low excitation foreground gas the amount of \cii calculated to arise from the \N+ region using the \nii as a tracer depends on the  \C+ to \N+ ratio.  It can be smaller than the value adopted from \cite{Goldsmith2015}, either due to differences in elemental abundances or to atomic processes that alter the ratio of $x$(C$^+$)/$x$(N$^+$) \citep{Langer2015X}.  For example, enhanced ionization by X-rays of \C+ to high ionization states (C$^{++}$, etc.) relative to \N+ multiple ionization (N$^{++}$, etc.), will create a deficit in \C+ compared to \N+. Multiply ionized states of C, N, and O are predicted in models of  \hii--PDR regions in the ionized boundary layers of clouds \citep{Abel2005} and in models incorporating large X-ray fluxes \citep{Langer2015X}.  In addition, the fine-structure lines of  [N\,{\sc iii}] at 57.3 \micron and  [O\,{\sc iii}] at 88.4 and 57.4 \micronno, respectively, have been observed in \hii regions, as well as Helium recombination lines, so some process is available to produce highly ionized carbon and nitrogen.  Unfortunately, [C\,{\sc iii}] and [C\,{\sc iv}] do not have fine--structure transitions, so we do not have a direct way to measure the multiple ionization of \C+.  For enhanced ionization of \C+ relative to \N+ to explain some (or most) of the increased \nii to \cii ratio, there must be a selective process that preferentially ionizes \C+ more efficiently than \N+ and/or a significant difference in recombination rates.  Thermal ionization by electrons can also enhance the loss of \C+ with respect to \N+ because of the different ionization potentials of carbon and nitrogen, but it requires high temperatures ($>$ 40,000 K).  X-rays or extreme ultraviolet (EUV) photoionization are possible explanations, however it will depend on the details of the relative cross sections, ionization rates, recombination rates, and gas temperatures. Such detailed modeling is beyond the scope of this paper.



\section{Summary}
\label{sec:summary}
To study the relative contribution of ionized and neutral regions of the ISM to the observed \cii emission we  identified 31 kinematic  \nii components with high signal to noise ($\ge$5) along ten lines of sight in the Galactic plane using spectrally resolved \cii (158 \micronno)   emission from the GOT C+ survey \citep{Langer2010,Pineda2013} and spectrally resolved \nii (205 \micronno) from \cite{Goldsmith2015}, both obtained with {\it Herschel} HIFI.  We used the \nii spectra to estimate how much \cii emission arises from the highly ionized gas and derive how much from the neutral gas containing only \C+ and atomic nitrogen, N$^o$.  Comparison of \nii and \cii show three important characteristics: 1) the ratio of \cii to \nii varies significantly from source to source and across the spectra, 2) the \nii and \cii lines are broad compared to those, such as $^{13}$CO, from the shielded regions of dense molecular clouds, and, 3) many  of the \cii lines are absorbed, sometimes strongly so, by foreground material and do not reflect the intrinsic source luminosity.  

The  \cii and \nii intensities  are used  to separate the amount of \cii arising from highly ionized gas versus that which arises from \C+ in the largely neutral gas.  However in a large number of components the results were unphysical, in that the   \cii calculated to be produced by the \nii region was relatively large and, in some cases, exceeded the observed \cii intensity.   Examination of the \cii and \nii spectra in Figures~\ref{fig:fig1_CII_NII_13CO} to \ref{fig:fig3_CII_NII_13CO} show that many \cii spectra appear to be absorbed and some strongly absorbed.  This absorption is likely responsible for some, if not all, of the modified observed ratio because the intrinsic \cii source intensity  will be larger than observed and thus the ratio of \cii from the ionized gas (estimated from \niino) to the total \cii will be reduced. In four sources we were able to calculate the effects of absorption on the \cii intrinsic intensity and showed that, by correcting for the foreground opacity, the fraction of \cii arising from the ionized gas with respect to the total \cii intensity is reduced by about 40\%. Three of these four sources resulted in fractional contributions from the \nii region, $I_{ion}$(\ciino)/$I_{tot}$(\ciino) about 35 to 60 percent, however one source was only reduced to 137 percent, which is unphysical.  

In summary our observations of spectrally resolved \cii and \niino, taken at face value, indicate that the contribution of \cii from \nii regions is large.  However, there is evidence that \cii spectra suffer from significant absorption and that one must be cautious in using \cii from ionized regions because of absorption of \cii by local foreground material or a foreground spiral arm.  We showed that it is possible in some cases to estimate the \cii absorption and calculate the intrinsic  \cii source intensity.  However, even with these corrections the fraction of \cii arising from \nii regions is large, raising the possibility that high density ionized gas is responsible for significant fraction of \cii emission.  Other factors that could change the interpretation of the fraction of \cii that arises from \nii gas include: 1) if the \cii lines are optically thick, and 2) if the fractional abundance ratio of \C+ to N$^+$ is different than assumed here, due either  to an intrinsic difference in abundance or ionization processes that preferentially ionizes \C+ with respect to N$^+$.  

Our analysis of the fractional contribution of \cii from \nii regions using spectrally resolved HIFI spectra are consistent with  the PACS \nii results showing that over 50\% of the lines of sight have large contributions of the ionized gas to the observed \ciino.  One must be especially careful in interpreting the properties of ionized and neutral regions with spectrally unresolved, or poorly resolved, \nii and \cii spectra, because the effects of \cii absorption can be large.  Finally, we note that the ten lines of sight observed with HIFI are among the strongest \cii sources in the GOT C+ survey, so we need further high spectral resolution observations of \nii and \cii from weaker sources to understand whether significant amounts of \cii arise from highly ionized regions along such lines of sight, and whether absorption is important in these regions too.




\begin{acknowledgements}
We thank Dr. T. Velusamy for providing the Galactic center \cii spectrum and for critical comments.   We also thank an anonymous referee for comments and suggestions that improved the discussion.  This work was performed at the Jet Propulsion Laboratory, California Institute of Technology, under contract with the National Aeronautics and Space Administration.   
 
\end{acknowledgements}



\bibliographystyle{aa}
\bibliography{aa_CII_NII_Langer_refs}
 

\end{document}